# A Unified DINOv2-Based Framework for LVEF Estimation, GLS Dysfunction Classification, and Early Cardiotoxicity Prediction

Xiaotong Zhang*[0000-0001-6085-2844], Mingyue Cui, Qing Cao, Jingming Xia

Cardiovascular Ultrasound R&D, GE HealthCare, Wuxi, China
* Corresponding author email: xiaotong.zhang@gehealthcare.com

**Abstract.** Left ventricular ejection fraction (LVEF) estimation (Task 1), global longitudinal strain (GLS)-based dysfunction classification (Task 2), and early cardiotoxicity prediction (Task 3) provide complementary information for cardio-oncology assessment. LVEF reflects macroscopic ventricular volume changes as the clinical standard, whereas GLS captures subtle myocardial deformation, indicating subclinical cardiotoxicity before overt LVEF decline. Furthermore, predicting cardiotoxicity from baseline echocardiography prior to treatment enables preventive interventions at an early stage. To address these three tasks, we employ a DINOv2-based framework with task-specific adaptation and prediction heads. Built upon a frozen foundation encoder, the framework incorporates parameter-efficient Low-Rank Adaptation (LoRA) and temporal aggregation to learn task-specialized representations, ensuring robust generalization. Crucially, during inference, it operates in a fully cycle-detection-free and phase-free manner, requiring neither cardiac cycle segmentation nor explicit End-Diastolic/End-Systolic (ED/ES) annotations. Additionally, we introduce an ED/ES-guided 2D/3D hybrid multi-view regression model specifically to optimize Task 1. On a patient-level split containing 1,203 training videos from 237 patients and 300 validation videos from 59 independent patients, the DINOv2-based framework achieved a mean absolute error (MAE) of 5.03% for Task 1, an AUC-ROC of 76.48% for Task 2, and an AUC-ROC of 70.26% for Task 3. For Task 1, the specialized ED/ES-guided model further improves performance, achieving an MAE of 4.64%. This framework demonstrates the effectiveness of foundation model representations across diverse cardio-oncology tasks and the additional benefit of physiology-guided modeling for accurate LVEF estimation. Our implementation is available at https://github.com/ZhangXiaotong015/EchoRisk-DINOv2.

**Keywords:** Cardio-Oncology, Echocardiography, Foundation Models.

## 1 Introduction

Heart failure (HF) is a progressive syndrome characterized by continuous cardiac remodeling [1][2]. In cardio-oncology, left ventricular ejection fraction (LVEF) and global longitudinal strain (GLS) are complementary echocardiographic biomarkers for monitoring cardiac function. While LVEF remains the most widely used metric, its quantification depends on accurate endocardial border delineation and is affected by image quality and measurement variability [3][4][5]. In contrast, GLS is more sensitive to subclinical dysfunction and early chemotherapy-related myocardial injury [6][7][8]. Furthermore, identifying patients at risk of treatment-related cardiotoxicity from baseline echocardiography may enable preventive interventions before irreversible damage occurs. Therefore, accurate assessment of LVEF, GLS, and cardiotoxicity risk is crucial for comprehensive cardio-oncology surveillance and risk stratification.

In clinical practice, automated quantitative echocardiography has transitioned to deep learning pipelines. The EchoNet-Dynamic framework establishes a representative benchmark for video-based LVEF estimation by utilizing a spatiotemporal 3D-CNN to directly regress cardiac function from echocardiography videos [9]. Deep learning models have also been used for echocardiographic view recognition and downstream functional analysis, demonstrating the feasibility of convolutional representation learning in echocardiography [10]. For functional video modeling, Two-Stream methods and CNN-RNN architectures combining 2D spatial encoders with temporal aggregators can be adapted to echocardiography-based EF estimation and abnormality classification [11]. However, benchmarks on Two-Stream and CNN-RNN architectures reveal that, although training loss may exhibit standard exponential decay, validation loss frequently becomes unstable or plateaus early under limited training data [11]. Models may also perform poorly at clinical extremes and show a tendency to regress toward the mean [11]. Therefore, standard video aggregation paradigms can suffer from representation smoothing, limiting their ability to preserve subtle motion cues in continuous, ungated, multi-cycle echocardiography videos. These limitations highlight the need for more transferable and robust representation learning strategies under limited clinical supervision.

Automated LVEF estimation, GLS-based LV dysfunction classification, and early cardiotoxicity prediction all rely on robust echocardiographic representations capable of capturing both global cardiac structure and subtle myocardial motion patterns. Under limited clinical data, fully trainable video encoders can easily overfit and produce overly smoothed representations. DINOv2 learns robust visual features without dense supervision and exhibits strong transferability across downstream visual tasks [12]. Low-Rank Adaptation (LoRA) enables parameter-efficient specialization of frozen foundation models through lightweight task-specific adaptation pathways [13]. Therefore, we employ a DINOv2-based foundation framework with task-specific adaptation and temporal aggregation across Task 1, Task 2, and Task 3. This design leverages a shared self-supervised visual backbone to support LVEF estimation, GLS-based dysfunction classification, and early cardiotoxicity prediction while reducing overfitting and improving generalization across heterogeneous cardio-oncology tasks despite limited clinical supervision.

Although the foundation-model framework provides a unified solution across all three tasks, LVEF estimation possesses a unique physiological characteristic. LVEF is determined by the difference between ED and ES states, making ED/ES information a natural inductive bias. Prior studies support this direction from multiple perspectives, showing that ED/ES cues improve EF prediction through joint phase-EF modeling [14], hybrid spatiotemporal architectures[15], , and frame-importance learning [16]. Cardiac phase detection methods also demonstrate that ED/ES phases can be automatically approximated without segmentation[17][18][19]. Multi-stream echocardiography studies also suggest that complementary views enhance robustness [20]. These motivate an additional ED/ES-guided 2D/3D multi-view regression model for Task 1.

In summary, our work makes three main contributions. First, we propose a unified DINOv2-based foundation-model framework for Task 1 LVEF estimation, Task 2 GLS-based LV dysfunction classification, and Task 3 early cardiotoxicity prediction. Second, we develop task-specific LoRA adaptation and phase-aware temporal modeling strategies that enable effective foundation-model representation learning across heterogeneous cardio-oncology objectives while maintaining cycle-free and phase-free inference. Third, we introduce an additional ED/ES-guided 2D/3D hybrid multi-view regression model for Task 1, allowing direct comparison between physiology-guided cardiac modeling and foundation-model-based LVEF estimation.

## 2 Methodology

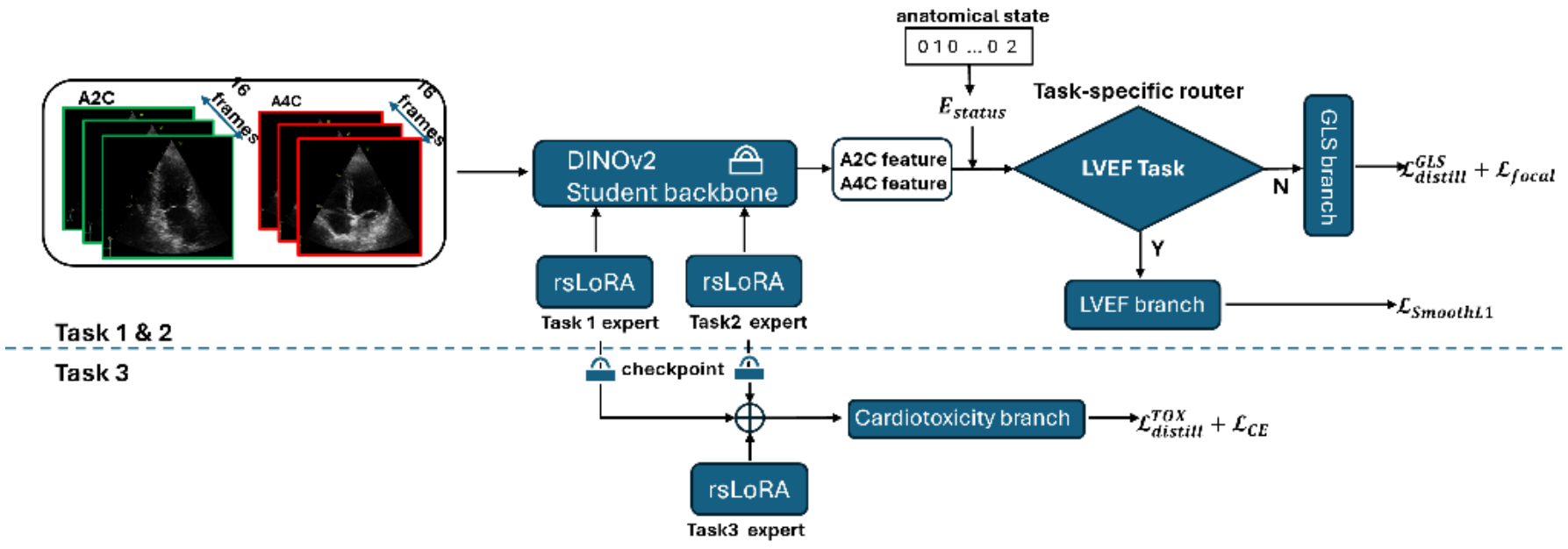

Fig. 1 Overview of the Proposed Multi-Expert DINOv2 Framework.

We develop two complementary modeling strategies for the EchoRisk challenge. The first is a unified DINOv2-based framework, as shown in Fig. 1, that addresses Task 1, Task 2, and Task 3 through parameter-efficient adaptation and task-specific prediction heads. The second is an ED/ES-guided 2D/3D hybrid regression model designed specifically for Task 1 LVEF estimation, serving as a physiology-guided alternative to foundation-model-based representations. Sections 2.1–2.4 describe the proposed DINOv2-based framework, including foundation-model feature extraction, task-specific adaptation, temporal modeling, and multi-task optimization. Section 2.5 presents the ED/ES-guided 2D/3D hybrid regression model, which serves as an additional physiology-guided approach specifically developed for Task 1 LVEF estimation. It should be noted that the ED/ES-guided 2D/3D hybrid regression model is not part of the DINOv2 pipeline but rather an independent method developed for Task 1.

### 2.1 Spatial feature extraction via foundational vision transformer

Given a multi-view echocardiography video sequences denoted as $X_{a2c}, X_{a4c} \in \mathbb{R}^{B \times M \times T \times C \times H \times W}$, where $B$, $M$, $T$, $C$, $H$, W represent the batch size, the number of cardiac cycles, the frame count per cycle, the image channel, the height, and the width respectively, we flatten the temporal dimensions for each single view independently to perform frame-level spatial embedding. The total frame sequence length for each view is as $K = M \times T$.

To leverage powerful visual priors without substantial training overhead, we employ a pre-trained DINOv2 [12] foundation model as the spatial backbone. For each constituent frame $x_k$, the spatial feature representation is derived by extracting the [CLS] token from the intermediate tensor of the final attention layer:

$$f_k = Extract_{CLS}\left(Layer_{last}\left(Backbone(x_k)\right)\right) \in R^D \quad (1)$$

where $D$ denotes the embedding channel dimension. This extraction yields two robust and stable frame-level feature sequences, denoted as $V_{raw_a2c}, V_{raw_a4c} \in \mathbb{R}^{B \times K \times D}$, preserving the independent anatomical semantics of each cardiac view.

### 2.2 Multi-expert routing via rank-stabilized LoRA

Simultaneous optimization of continuous LVEF regression and binary GLS dysfunction classification often induces cross-task gradient interference due to their distinct visual cues and learning objectives. To decouple task-specific representations while maintaining parameter efficiency, we introduce a task-specialized expert routing mechanism based on rank-stabilized LoRA (rsLoRA).

For each Vision Transformer (ViT) block, the query-key-value (QKV) projection is augmented with task-specific low-rank adaptation branches. During the joint optimization stage, two independent expert adapters, denoted as $\Phi_{lvef}$ and $\Phi_{gls}$, are assigned to LVEF regression and GLS-based dysfunction classification respectively. Instead of standard LoRA, rsLoRA [21] is adopted to improve optimization stability under low-rank parameterization:

$$W_{active} = W_0 + \frac{\alpha}{\sqrt{r}} B_{task} A_{task}, task \in \{gls, lvef\} \quad (2)$$

where $W_0$ denotes the frozen base weight, $r$ is the adapter rank, and $\alpha$ is a scaling factor. Matrix $A$ is initialized using Kaiming-uniform initialization [22], whereas matrix $B$is initialized to zero.

Building upon parameter-efficient adaptation techniques such as adapters and LoRA [25][13], a Dynamic Pointer Redirection (DPR) mechanism is employed to activate only the adapter corresponding to the current task during forward propagation. Consequently, inactive experts are excluded from gradient propagation, effectively isolating task-specific optimization while preserving the computational efficiency of the frozen DINOv2 backbone.

For Task 3, we adopt a sequential adaptation strategy. After joint optimization of the LVEF and GLS experts, the resulting checkpoint is used to initialize cardiotoxicity prediction. A third rsLoRA expert, Φtox, is subsequently introduced and optimized on the cardiotoxicity cohort. During training, the previously learned LVEF and GLS experts remain frozen and act as fixed feature extractors, providing complementary macro-volumetric and micro-deformation priors. Their representations are combined with the cardiotoxicity-specific features learned by Φtox through late feature fusion, enabling knowledge transfer from cardiac function assessment tasks to early cardiotoxicity prediction.

### 2.3 Phase-aware temporal synchronization encoding

To incorporate clinically meaningful cardiac phase information into sequential modeling, we introduce a phase-aware temporal synchronization encoding module. Each sampled frame is assigned a lightweight anatomical state label:

$$s_k = \begin{cases} 1, & ED\ frame \\ 2, & ES\ frame \\ 0, & otherwise \end{cases} \tag{3}$$

where ED and ES denote end-diastolic and end-systolic cardiac phases, respectively.

The state variable is projected through a learnable embedding matrix $E_{status} \in \mathbb{R}^{3\times D}$ to obtain a phase-aware synchronization vector, which is fused with the corresponding visual token before temporal aggregation.

A key characteristic of the proposed design is its asymmetric training–inference paradigm. During training, the model is exposed to both phase-aligned and blind-sampled temporal sequences, enabling it to exploit sparse phase supervision while remaining robust to imperfect annotations. During inference, explicit phase information is unavailable and all frames are assigned the neutral state ($s_k \equiv 0$). Consequently, the model must infer cardiac dynamics directly from image content, avoiding any dependence on external phase annotations.

### 2.4 Multi-Task Joint Optimization with Asymmetric Distillation

To jointly optimize GLS classification and LVEF regression, we employ two task-specific experts with independent temporal encoders and prediction heads. The GLS expert focuses on learning myocardial deformation patterns, whereas the LVEF expert models global ventricular dynamics.

Given the limited training data and the higher susceptibility of binary classification to overfitting, knowledge distillation is applied to the GLS branch. A frozen teacher network (DINOv2 ViT-B/14) provides auxiliary supervision through Kullback–Leibler (KL) divergence, encouraging alignment between student and teacher feature distributions [24]. The GLS expert is optimized using a focal classification loss combined with the distillation objective. In contrast, the LVEF branch is optimized solely through Smooth L1 regression loss without distillation, preserving the flexibility of ventricular function representations.

The objective for the joint Task 1–Task 2 training stage is defined as

$$\mathcal{L}_{Task1/Task2} = \lambda_l \mathcal{L}_{SmoothL1} + \lambda_d \mathcal{L}_{distill}^{GLS} + \lambda_g \mathcal{L}_{focal} \tag{4}$$

where $\mathcal{L}_{distill}^{GLS}$ denotes the KL-based distillation loss, $\mathcal{L}_{focal}$ is the dysfunction classification loss, and $\mathcal{L}_{SmoothL1}$ is the LVEF regression loss.

For Task 3, During optimization, the previously learned GLS and LVEF experts remain frozen and act as fixed feature extractors, providing complementary micro-deformation and macro-volumetric priors. Their representations are fused with cardiotoxicity-specific features learned by the Task 3 expert through a late-fusion prediction module. Similar to Task 2, knowledge distillation is applied to preserve semantic consistency with the teacher representation, while cardiotoxicity prediction is supervised using a cross-entropy classification loss. The corresponding objective is:

$$\mathcal{L}_{task3} = \lambda_t \mathcal{L}_{distill}^{TOX} + \lambda_c \mathcal{L}_{CE} \quad (5)$$

### 2.5 ED/ES-Guided 2D/3D Hybrid LVEF Estimation

An ED/ES-guided hybrid regressor is proposed to estimate LVEF by integrating full-cycle video dynamics, key-phase structural differences, and multi-view information. Each examination is formatted as a two-slot input$X = \{X_{A4C}, X_{A2C}, m_{A4C}, m_{A2C}\}$, where $X_v \in \mathbb{R}^{B\times1\times T\times H\times W}$ denotes the cine loop (zero-filled if absent) and $m_v \in \{0,1\}$ indicates view availability.

The first branch captures global systolic-diastolic motion. The available cine loops are uniformly sampled and processed via a shared 3D video encoder $F_{3D}$ (e.g., R(2+1)D) and a projection head $P_{full}$ to yield full-video embeddings:

$$z_{full}^{v} = P_{full}(F_{3D}(X_v)) \quad (6)$$

The second branch explicitly encodes contraction-related structural changes. Given the ED/ES phase indices $i_{ED}, i_{ES}$ from the preprocessing module, a shared 2D encoder $F_{2D}$ extracts frame-level features $f_{ED}^{v}$and $f_{ES}^{v}$. To introduce physiological priors without manual segmentations, the original, signed, and absolute difference features are concatenated and projected via $P_{EDES}$:

$$z_{EDES}^{v} = P_{EDES}([f_{ED}^{v}, f_{ED}^{v}, f_{ED}^{v} - f_{ED}^{v}, |f_{ED}^{v} - f_{ED}^{v}|]) \quad (7)$$

For each view, a sample-wise gate$g^v = \sigma\left(G\left(\left[z_{full}^{v}, z_{EDES}^{v}\right]\right)\right)$ adaptively balances the two branches, yielding the fused view embedding:

$$z^v = g^v \odot z_{full}^{v} + (1 - g^v) \odot z_{EDES}^{v} \quad (8)$$

The final exam-level LVEF prediction is generated by a missing-view-aware fusion head H: $\hat{y} = H([m_{A4C}Z^{A4C}, m_{A2C}Z^{A2C}, m_{A4C}m_{A2C}])$.

The model is optimized via a multi-task loss combining exam-level regression, view-level auxiliary supervision, and cross-view consistency regularization:

$$\mathcal{L} = Huber(\hat{y}, y) + \lambda_v \sum_{v\in\{A4C,A2C\}} m_v Huber(\hat{y}, y)$$
$$+\lambda_c m_{A4C} m_{A2C} SmoothL1(\hat{y}_{A4C}, \hat{y}_{A2C}) \quad (9)$$

The three terms optimize overall accuracy, enhance single-view robustness, and enforce clinical consistency between coexisting views, respectively.

# 3 Dataset and pre-processing

## 3.1 Dataset

We evaluated our approach on the EchoRisk dataset released for the MICCAI EchoRisk 2026 Challenge [23]. The dataset originates from the CARDIOCARE prospective multicenter study and includes echocardiographic examinations acquired from five European clinical centers. Videos were collected using multi-vendor ultrasound systems under routine clinical conditions and include apical four-chamber (A4C) and apical two-chamber (A2C) views.

The challenge defines three benchmark tasks: cardiac parameter estimation (Task 1), left ventricular dysfunction detection (Task 2), and early cardiotoxicity prediction (Task 3). Tasks 1 and 2 comprise 422 patients with 1,123 examinations and 2,159 echocardiography videos collected across longitudinal follow-up visits, whereas Task 3 includes 280 patients with baseline examinations and expert-adjudicated cardiotoxicity labels. Detailed information regarding data acquisition, label generation, and cohort statistics is provided in the official challenge description [23].

## 3.2 Video pre-processing

For each examination, end-diastolic (ED) and end-systolic (ES) landmarks were identified [19] and used to construct cycle-aware temporal representations. For the DINOv2-based framework, up to two cardiac cycles were retained per study, and 16 frames were uniformly sampled between the ED and ES positions for each cycle. To improve robustness to temporal localization errors and acquisition variability, ten temporally distinct sampling groups were generated offline. These groups included the original ED-ES sampling strategy, temporally perturbed variants, and blind temporal sampling across the entire video duration. During training, one sampling group was randomly selected as input to provide temporal augmentation, whereas during inference only blind temporal sampling was adopted. All frames were resized to 224×224 pixels and replicated to three channels before being fed into the network. For the ED/ES-guided hybrid LVEF estimator, all training videos were used; each DICOM cine loop was normalized, cropped to the nonzero region, resized to 160×160 pixels, and sampled into 64-frame grayscale clips.

# 4 Experiments and results

## 4.1 Implementation details

The DINOv2-based framework was initialized from the official DINOv2 ViT-B/14 distilled checkpoint [12]. rsLoRA adapters with rank $r = 4$ and scaling factor $\alpha = 1.0$ were injected into the frozen backbone. Optimization was performed using AdamW with a base learning rate of $1\times10^{-4}$; LoRA parameters and newly introduced trainable modules used learning-rate scaling factors of 0.1 and 1.0, respectively. The loss weights were set to (λl, λd, λg) = (2.0, 1.0, 1.0) for joint Task 1–Task 2 training and (λt, λc) = (1.0, 1.0) for Task 3. Each batch contained two examinations with 16 A2C and 16 A4C frames, and gradients were accumulated over 16 iterations. Tasks 1 and 2 were jointly trained for 100 epochs with 100 iterations per epoch. To enhance representation learning, the GLS branch employed a 16-channel classification head during joint training. Task 3 was subsequently trained for 20 epochs with 32 iterations per epoch, initialized from the checkpoint obtained from the jointly trained Task 1-Task 2 model using the 2-channel classification-head configuration. Decision thresholds of 0.32 (Task 2) and 0.008 (Task 3) were used to map predicted probabilities to binary predictions. The low Task 3 threshold likely reflects under-confident probability estimates resulting from adapting DINOv2 to a limited cardiotoxicity dataset, highlighting a calibration limitation of the current approach. While threshold-dependent metrics such as balanced accuracy varied with the selected threshold, AUC-ROC remained unchanged because it depends on the ranking of predictions rather than the specific decision boundary. A fixed random seed of 42 was used throughout both training and inference for all experiments. For the ED/ES-guided hybrid LVEF estimator, all training videos were used, and the model was trained for 40 epochs using AdamW with a learning rate of $2\times10^{-4}$, batch size 2, mixed precision, Huber loss, an auxiliary view loss weight of 0.25, and a view-consistency loss weight of 0.05. All reported results correspond to the checkpoint achieving the best primary validation metric. All experiments were conducted on a single NVIDIA RTX PRO 5000 Blackwell GPU with 48 GB of memory.

As described in Section 3.2, the DINOv2 pipeline was trained and validated using ten offline-extracted frame sets generated from each video. Validation was performed using a randomly selected frame set, and the results reported in Table 1 were obtained under this offline extraction protocol. However, in the official test environment, frames were extracted online directly from the input DICOM files within the Docker container. To better match this deployment scenario, model selection for final test-time inference was based on validation performance obtained under the online frame-extraction setting. While a full component ablation study is beyond the scope of this work, empirical observations from our experiments suggest that rsLoRA-based adaptation contributed the largest performance gain, while phase-aware temporal encoding and asymmetric distillation provided additional complementary improvements.

### 4.2 Quantitative comparison with baseline

Table 1. Quantitative comparison with the challenge baseline on the validation set.

| Task | Model | Metric | Val [23] (Baseline) | Val (Ours) |
|---|---|---|---|---|
| Task 1 | DINOv2 pipeline | MAE (pp) ↓ | 5.02% | 5.03% |
| | ED/ES-guided model | | | 4.64% |
| Task 2 | DINOv2 pipeline | AUC ↑ | 75.50% | 76.48% |
| Task 3 | DINOv2 pipeline | AUC ↑ | 64.60% | 70.26% |

Table 1 summarizes the validation performance of the proposed methods and the challenge baseline under a fixed random seed of 42. For Task 1, the DINOv2-based framework achieved a comparable MAE of 5.03%, while the proposed ED/ES-guided 2D/3D hybrid model further reduced the error to 4.64%, demonstrating the benefit of incorporating physiologically meaningful ED/ES information for LVEF estimation. For Task 2, the proposed DINOv2 framework improved the AUC from 75.50% to 76.48%, indicating enhanced discrimination of GLS-based LV dysfunction. The largest gain was observed for Task 3, where the AUC increased from 64.60% to 70.26%, suggesting that the proposed sequential adaptation and expert-fusion strategy effectively transfers cardiac function knowledge to early cardiotoxicity prediction.

To ensure the reliability of the evaluation results on the validation dataset, we also reran our methods for all tasks using seeds 44, 46 and 48, in addition to seed 42 we reported in Table 1. Mean values and standard deviations are reported across seeds. For Task 1, $mean \pm std$ is 5.07$\pm$0.06 pp and 4.70 $\pm$ 0.26 pp for our DINOv2-based and ED/ES guided methods, respectively. For Task 2 and Task 3, the $mean \pm std$ values for the DINOv2 method across seeds are 75.48 $\pm$ 1.93% and 67.04%$\pm$2.62%, respectively. These evaluations indicate that our DINOv2 method is stable across all three tasks and seeds. Furthermore, we note that while both the Task 1/2 dataset and the Task 3 dataset were independently partitioned into training and validation sets, an overlap exists between them; statistics show that 40 videos from the training dataset of Task 1/2 appear in the validation set of Task 3. Although the training targets for Task 1/2 and Task 3 differ, since the weights pre-trained on Task 1/2 were utilized as the feature extractor for Task 3, a certain degree of validation data leakage may occur, which could potentially impact the final AUC results of Task 3.

Overall, the results demonstrate the effectiveness of foundation-model-based representations across diverse cardio-oncology tasks, with additional physiological modeling providing further gains for LVEF estimation.

## 5 Conclusion

We developed a unified DINOv2-based framework for the three EchoRisk tasks and an additional ED/ES-guided model for Task 1 LVEF estimation. The experimental results demonstrate the effectiveness of foundation-model representations across diverse cardio-oncology tasks and the added value of physiology-guided modeling for accurate LVEF assessment.

**Compliance with Ethical Standards:** This study utilizes the fully de-identified, publicly available dataset provided by the EchoRisk-MICCAI Challenge organizers. The data collection and primary ethical approvals were managed by the challenge host institutions. No additional ethical approval or administrative permission was required from the authors for this retrospective analysis.

**Funding and Disclosure of Interests:** All authors are current employees of GE Healthcare. This study received no specific grant or extra funding from any funding agency in the public, commercial, or not-for-profit sectors. Standard operational support, including computer hardware and registration fees, was provided by GE Healthcare. The authors declare no other competing financial or non-financial interests.